\documentclass[12pt]{article}

\usepackage{latexsym,amssymb,euscript}
\usepackage[dvips]{graphicx}
\usepackage{amsmath}
\usepackage{graphbox}
\usepackage{amsfonts}
\usepackage{amssymb, epsfig} 

\usepackage{float}
\usepackage{array}

\usepackage[dvipsnames]{xcolor}

\usepackage[utf8]{inputenc}

\usepackage{color,soul}

\begin{document}

\begin{center}
{\Large \textbf{Confinement-Higgs and deconfinement-Higgs transitions in
three-dimensional $Z(2)$ LGT}}

\vspace*{0.6cm}
\textbf{B. All\'es\footnote{email: alles@pi.infn.it}} \\
\vspace*{0.1cm}
\centerline{\it INFN Sezione di Pisa, Largo Pontecorvo 3, 56127 Pisa, Italy}
\vspace*{0.3cm}
\textbf{O.~Borisenko\footnote{email: oleg@bitp.kyiv.ua}} \\
\vspace*{0.1cm}
\centerline{\it INFN Gruppo Collegato di Cosenza, Arcavacata di Rende, 87036 Cosenza, Italy}
\centerline{\rm and}
\centerline{\it N.N.Bogolyubov Institute for Theoretical Physics,}
\centerline{\it National Academy of Sciences of Ukraine, 03143 Kiev, Ukraine}
\vspace*{0.3cm}
\textbf{A. Papa\footnote{email: papa@fis.unical.it}} \\
\vspace*{0.1cm}
\centerline{\it Dipartimento di Fisica, Universit\`a della Calabria}
\centerline{\rm and}
\centerline{\it INFN Gruppo Collegato di Cosenza, Arcavacata di Rende, 87036 Cosenza, Italy}
\end{center}


\begin{abstract}
We re-examine by numerical simulation the phase structure of the three-dimensional Abelian lattice gauge theory (LGT) with
$Z(2)$ gauge fields coupled to $Z(2)$-valued Higgs fields. Concretely, we explore two different order parameters which are able to distinguish the three
phases of the theory: (i) the Fredenhagen-Marcu operator used to discriminate between deconfinement and confinement/Higgs phases and (ii) the Greensite-Matsuyama overlap
operator proposed recently to distinguish confinement and Higgs phases. The latter operator is an analog of the overlap Edwards-Anderson order parameter for spin-glasses.
According to it, the Higgs phase is realized as a glassy phase of the gauge system. For this reason standard tricks for simulations of spin-glass phases are utilized in this
work, namely tempered Monte Carlo and averaging over replicas. In addition, we also present results for a certain definition of distance between Higgs field configurations.
Finally, we calculate various gauge-invariant correlation functions in order to extract the corresponding masses. 
\end{abstract}

\section{Introduction}

\begin{figure}[htb]
\centering
\includegraphics[width=0.65\linewidth,clip]{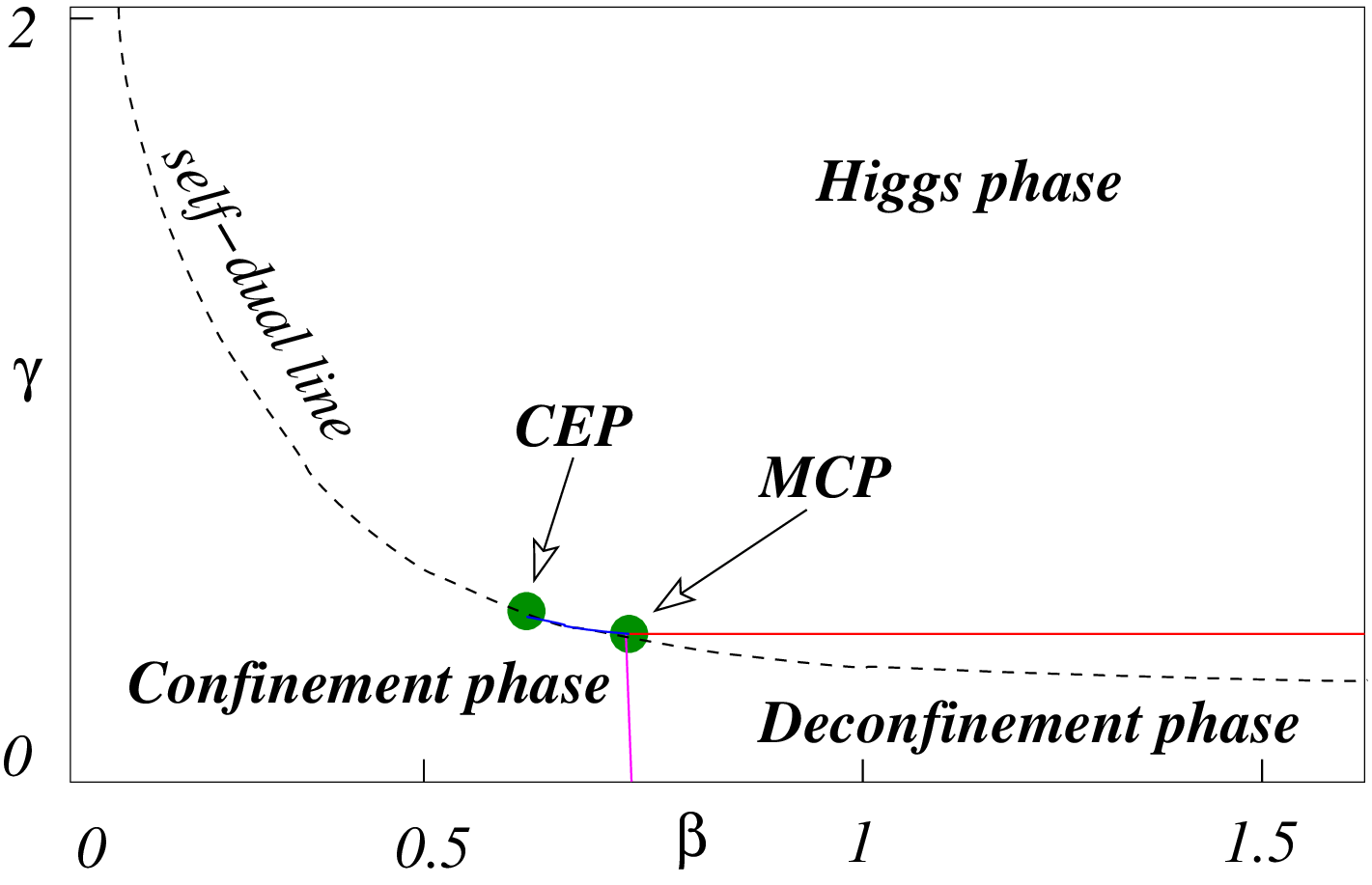}
\caption{Phase diagram of the $Z(2)$ gauge-Higgs LGT. See text for details.}
\label{fig:phase_diagram_general}
\end{figure}

In this paper we shall study the three-dimensional ($3d$) $Z(2)$ gauge-Higgs theory on the lattice. The partition function of the model on $\Lambda\in Z^3$ is defined as 
\begin{equation}
\label{gauge_higgs_pf}
Z_{\Lambda}(\beta, \gamma) =  
\sum_{\{ z_n(x)=\pm 1 \} } \ \sum_{\{ s(x)=\pm 1 \} } \ e^{S_G + S_H} 
\end{equation} 
with the gauge $S_G$ and the Higgs $S_H$ actions given by 
\begin{eqnarray}
\label{gauge_action} 
S_G &=& \beta \sum_{x,n<m} z_n(x) z_m(x+e_n) z_n(x+e_m) z_m(x)  \ ,   \\ 
\label{higgs_action}   
S_H &=& \gamma \sum_{x}\sum_{n=1}^3  \ s(x) z_n(x) s(x+e_n) \ . 
\end{eqnarray} 
Here, $x=(x_1,x_2,x_3)$ with $x_n\in [0,L-1]$ denotes a site on a lattice of linear size $L$ and $l=(x,n)$ denotes a link pointing
from the site $x$ in the direction $n$. Units are chosen to have the lattice spacing $a=1$. Periodic boundary conditions are also imposed in all directions.
The model is invariant under the action of local $Z_{\rm l}(2)$ transformations $\omega(x)$ and global  $Z_{\rm gl}(2)$ transformation $\alpha$:
\begin{eqnarray}
\label{zn_transform}
z_n(x) &\rightarrow& z_n^{\prime}(x) = \omega(x) z_n(x) \omega(x+e_n) \ ,  \\ 
s(x) &\rightarrow& s^{\prime}(x) = \omega(x) s(x) \alpha  \ .
\end{eqnarray}
Thus the full symmetry group is $Z_{\rm l}(2)\times Z_{\rm gl}(2)$. This model is self-dual in that
up to a constant, the partition function on the dual lattice maintains the form shown in Eq.(\ref{gauge_higgs_pf}) upon replacing the couplings $\beta,\gamma$ for the dual couplings
\begin{eqnarray}
\label{3d_dual_coupl}
\gamma \rightarrow \gamma_d = -\frac{1}{2} \ln \tanh \beta \ ,  \qquad
\beta \rightarrow \beta_d = -\frac{1}{2} \ln \tanh \gamma \ .
\end{eqnarray}
The study of this model has a long history and the general form of its phase diagram is well known \cite{drouffe_75,fradkin_shenker,creutz_80,stack_80}. Fig.\ref{fig:phase_diagram_general} exhibits a schematic summary of this diagram.
Along the $\gamma=0$ line the pure gauge model is dual to the $3d$ Ising model, whose second order critical point is known with high accuracy, $\beta_{\rm c}\approx 0.76141$. In the $\beta\to\infty$ limit the model is again reduced to the
$3d$ Ising model with critical coupling $\gamma_{\rm c}\approx 0.22165$. Two critical lines, dual to each other, emanate from those critical points and enter the $(\beta, \gamma)$-plane. They are the magenta and red lines in Fig.\ref{fig:phase_diagram_general} and correspond to the second order phase transition
in the Ising universality class \cite{on_gauge_z2}. These two lines eventually meet at a self-dual point called multi-critical point (MCP) at $\beta\approx 0.7525$, $\gamma\approx 0.2258$. 
The scaling behavior in the MCP is governed by the $3d$ $XY$ universality class \cite{z2_gauge-higgs_mcp}. Over the short interval between MCP and the critical end point (CEP) along the self-dual line,
the critical line turns into a first order line \cite{torrero_03}. It terminates in the CEP at $\beta\approx 0.689$, $\gamma\approx 0.258$. The scaling continuum limit can be described by the $\Phi^4$
theory\footnote{The results of \cite{brezin_82} were obtained by mean-field approximation. We are not aware of any numerical confirmation of these results.} \cite{brezin_82}.
No conventional critical behavior was found on the left side of the CEP. This is of course a consequence of the Fradkin-Shenker \cite{fradkin_shenker} and Osterwalder-Seiler \cite{seiler_78} theorems.

 As is evident from above description, the phase diagram contains three phases. At the right bottom corner there is a deconfinement phase
 separated by thermodynamic transitions from the confinement and the Higgs phases.

 Two order parameters can distinguish deconfinement from confinement and Higgs phases. The first of these order parameters is the Fredenhagen-Marcu (FM) operator introduced in Ref.\cite{FM_86}
 and tested in Monte Carlo  simulations in \cite{FM_86_MC} on the $4d$ $Z(2)$ gauge-Higgs theory, which has a similar phase structure. The rigorous proof that the FM operator behaves in the
 expected way was provided recently in Ref.\cite{FM_24}.
 The FM operator has been studied also in Ref.~\cite{toric_code_24}, in
    a closely related toric code model, which has a similar phase structure.

 The second order parameter is the Preskill-Krauss (PK) operator introduced in Ref.\cite{kraus2}.  For general $Z(N)$, the PK operator is based on the Aharonov-Bohm effect: it can detect $Z(N)$ charges in
 the deconfinement phase. The finite-temperature extension of the PK operator was constructed and studied in \cite{borisenko_lat97,borisenko_98}. 
 Essentially, both the FM and the PK operators are nonlocal and compare two screening mechanisms of discrete $Z(N)$ charges in the system: a dynamical one due to matter fields and a screening due to gauge fields. 
 Important in the context of the present paper is that both order parameters, FM and PK, cannot distinguish the confinement phase from the Higgs phase.

 According to conventional lore, no thermodynamic phase transition exists in the $\gamma < 0.258$ region of the phase diagram, {\it i.e.} below the CEP. 
 This lore originates from the Fradkin-Shenker result \cite{fradkin_shenker} on the existence of an analytic path between any two points in the confinement/Higgs region.
 The free energy and all its derivatives are analytic functions of the coupling constants along such a path. This result, however, does not rule out the presence of critical behavior.
 One possible way to detect that critical behavior is to fix an appropriate gauge as, according to the Elitzur theorem \cite{elitzur_75}, local gauge symmetry cannot break spontaneously,
 so that expectation values of the Higgs field vanish in the theory without gauge fixing. It was mentioned in Ref.\cite{grady_05} that fixing a noncomplete axial gauge such that it still
 leaves an unbroken $Z(2)$ global symmetry over a $2d$ plane leads to a nonzero value of the Higgs field above the self-dual line, on the left hand side of the CEP. That $Z_{\rm gl}(2)$ symmetry gets
 spontaneously broken in the Higgs phase. The physical significance of this result remains unclear because the Higgs expectation value does depend on the gauge.   
 
 Another, more promising, route to recognize critical behavior is to use gauge-invariant nonlocal order parameters. Neither the FM nor the PK operators can do the job. 
 Still, it is possible to construct such an operator using the close analogy between gauge-Higgs systems and the theory of spin-glasses. Indeed, the gauge-Higgs part
 (\ref{higgs_action}) of the action coincides with the Ising spin-glass  (Edwards-Anderson) model \cite{edwards_75}. Here, the gauge field plays the role of
 position-dependent coupling constant with a probability distribution weighted by the Wilson action. This analogy has been deeply explored in
 Refs.\cite{greensite_17_Sconf,greensite_18_overlap,greensite_20_overlap,greensite_22_overlap} for the $SU(2)$ gauge-Higgs action with scalar fields in the fundamental
 representation. Moreover, a gauge-invariant analog of the overlap operator, the Greensite-Matsuyama operator, for spin-glasses has been put forward in these papers. Monte Carlo computations of the proposed
 operator did reveal a nontrivial critical behavior which is attributed to the spontaneous symmetry breaking of the global $Z(2)$ group acting only on the Higgs fields
 (so-called custodial symmetry). Thus, a quantity able to distinguish confinement from Higgs phases exists. A physical distinction between the two phases has been also
 described in the above-mentioned papers and it relies upon a different realization of confinement: C-confinement in the Higgs phase (existence of color neutral states only)
 and S-confinement in the confinement region (existence of string tension at short and intermediate distances, and of string breaking at large distances).  In this picture the
 Higgs phase is realized as a glassy phase with nonzero value of the overlap operator, while the confinement phase is viewed as a paramagnetic phase with vanishing overlap operator. 
 
 In this paper we extend the just described picture to the $Z(2)$ gauge-Higgs model (\ref{gauge_higgs_pf}). For simplicity, we treat the $3d$ model at zero temperature. 
 On the one hand, as a theory with discrete degrees of freedom, this model is simpler than the $SU(2)$ one. On the other hand, there is a third phase characterized by deconfinement which
 is not present in the $SU(2)$ theory. From the spin-glass point of view, deconfinement can be considered as a ferromagnetic phase. If so, the Greensite-Matsuyama overlap operator should
 vanish also in this phase. We expect, therefore, that the overlap operator can distinguish both confinement and deconfinement phases from the Higgs phase, but not confinement from
 deconfinement. This also means that in order to give a full characterization of the phase diagram of the model we will have to measure both the Greensite-Matsuyama overlap
 operator and the FM and/or PK operators. This is what we will accomplish in the present study concentrating on the overlap and FM operators\footnote{We have measured the PK operator but, despite having gathered a big statistics, the error bars remained too large, a fact that prevented us from attaining reliable conclusions. An even bigger statistics and improved algorithms are needed to numerically evaluate the PK operator.}. 
 In addition, we will compute gauge-invariant correlation functions like
 plaquette-plaquette and correlation between Wilson lines of different geometries. These correlations will enable us to extract the masses of gauge and Higgs fields.
 Finally, we also compute the correlation between independent configurations of the Higgs fields on a fixed spatial slice.

 The paper is organized as follows.  In Sec.2 we define all operators involved in the analysis. These include the Greensite-Matsuyama overlap operator, the FM operator, correlation functions and distances in configuration space.  
 In Sec.3 we describe the Monte Carlo updating and present the results of our simulations. Sec.4 summarizes our results.

\section{Order parameters and correlation functions}
\label{observables}

\subsection{Overlap and related operators}

We have utilized the overlap operator defined in \cite{greensite_20_overlap}. It consists in performing a series of $n_{\rm sym}$ special Monte Carlo updatings on a copy of the running configuration
in equilibrium at the assigned values of $\beta,\gamma$. The word ``special'' has a twofold meaning: first, the updating must act on all gauge and Higgs fields except for the spatial links lying on a given slice at a fixed value of the
temporal coordinate. This temporal coordinate will usually be taken to be zero. Second, those updatings must be slow, that is, the successive configurations must not be too decorrelated.

Once the above $n_{\rm sym}$ special Monte Carlo updatings have been applied, the operator to be evaluated is
\begin{equation}
  G\equiv\frac{1}{V_{\rm space}}\sum_{\vec{x}}\left\vert\frac{1}{n_{\rm sym}}\sum_{{m}=1}^{n_{\rm sym}}s^{(m)}(0,\vec{x})\right\vert\;,
  \label{overlapoperator}
\end{equation}
where $s^{(m)}(0,\vec{x})$ is the $m$-th measurement of the Higgs field at spatial position ${\vec{x}}$ over the $t=0$ slice with $1\leq m\leq n_{\rm sym}$, and $V_{\rm space}=L^2$ is the spatial volume of the slice.

In principle $n_{\rm sym}$ can be tuned in order to choose the best estimator. But a better strategy consists in repeating the measurements of (\ref{overlapoperator}) for
a number of not-too-large values of $n_{\rm sym}$ and fit the results with the functional form
\begin{equation}
  G=\langle G\rangle+\frac{A}{\sqrt{n_{\rm sym}}}\;,
  \label{overlapoperator_fit}
\end{equation}
where $\langle{G}\rangle$ indicates the physically meaningful result and $A$ is a constant. The functional form (\ref{overlapoperator_fit}) is expected from general statistical arguments.

As stated above, a further important aspect of the measurement of (\ref{overlapoperator}) is that the Monte Carlo updatings must be slow. This is due to the fact
that the overlap operator measures a transient phenomenon. Thus, if a heavy decorrelation were applied (for instance by using a cluster algorithm or by repeating many
hits of more conventional local algorithms like Metropolis or Heat Bath), one would obtain that $\langle{G}\rangle$ in (\ref{overlapoperator_fit})
tends to zero for any value of the couplings $\beta,\gamma$.

Another useful observable is $r_{\alpha,\beta}$ which provides a measure of the distance between two configurations labelled $\alpha$ and $\beta$.
This observable
has been introduced in \cite{distance_config} and tested on $2d$ Ising, $Z(N)$ and $XY$ models. A possible extension to our gauge-Higgs $Z(2)$ model consists in simulating in parallel several replicas of the
fields and extracting the normalized quantity 
\begin{equation}
r_{\alpha, \beta} \equiv \frac{1}{2 V_{\rm space}}\sum_{\vec{x}} \frac{1}{n_{\rm sym}}\sum_{{m}=1}^{n_{\rm sym}} \left\vert s_{\alpha}^{(m)}(0,\vec{x}) - s_{\beta}^{(m)}(0,\vec{x}) \right\vert\;,
\label{distance_operator}
\end{equation}
where $\alpha,\beta$ labels two different replicas and, similarly as above, $s_{\alpha}^{(m)}(0,\vec{x})$ is the $m$-th measurement
of the $\alpha$th replica of Higgs fields over the temporal slice $t=0$. Again $m$ designates measurements, $1\leq m\leq n_{\rm sym}$.

\subsection{$FM$ order parameter}
\label{2.2FM}

Two order parameters have been proposed in the past to distinguish the deconfinement phase from the confinement and/or the Higgs phases.
In this paper we define and study the FM order parameter \cite{FM_86}. It is defined as follows. Let ${\cal{C}}$ be a closed rectangular loop of links with one side of length $T$
along the temporal coordinate and the other side of length $R$ along one spatial direction. We denote by $W({\cal{C}})$ the corresponding Wilson loop, 
\begin{equation}
\label{wilson_loop_def}
W({\cal{C}}) \equiv \prod_{l\in \cal{C}} \ z(l) \ . 
\end{equation}
By cutting $W({\cal C})$ into two halves along the temporal sides, we obtain a $\sqcap$ and a $\sqcup$ shaped objects. Let us take one of them, for instance
$\sqcap$, and call it ${\cal L}_{xy}$ where $x,y$ are the endpoints. Hence, the horizontal side of $\sqcap$ runs along a spatial direction and has length $R$ while the vertical sides run along the temporal direction and have lengths $T/2$. We then consider a gauge invariant Wilson line, denoted by $V({\cal L}_{xy})$, by multiplying ${\cal L}_{xy}$ with the Higgs fields at the endpoints,
\begin{equation}
\label{wilson_line_def}
V({\cal{L}}_{xy}) \equiv s(x) \ 
\Big( \prod_{l\in {\cal{L}}_{xy}} \ z(l) \Big) \ s(y) \ . 
\end{equation}
Given the $R$-dependent ratio
\begin{equation}
\label{FMoperator_def}
H(R) = \frac{\langle V({\cal{L}}_{xy}) \rangle^2}{\langle W({\cal{C}}) \rangle} \ ,
\end{equation}
the FM operator is defined as
\begin{equation}
\label{FMoperator_defBIS}
\rho \equiv \lim_{R\to\infty} \ H(R) \ .
\end{equation}

\subsection{Correlation functions} 
\label{2.3Corr}

Let $p$ and $p^\prime$ indicate plaquettes. The standard connected wall-wall plaquette correlation function is defined as 
\begin{eqnarray}
\label{plaq_plaq_corr}
P(R) \equiv \langle \sum_p z(p) \sum_{p^\prime}z(p^{\prime}) \rangle - 
 \langle z(p)  \rangle^2 \ ,
\end{eqnarray}
where $z(p)$ is the product of gauge links around the plaquette $p$ and the sums run over the plaquettes $p$ and $p^\prime$ placed on parallel planes (the walls) separated by a distance $R$.
It is expected that this correlation decays exponentially in the confinement and the Higgs phases and can be used to compute the mass of bounded states of gauge fields. 

Wall-wall correlations of Wilson lines are defined similarly as 
\begin{eqnarray}
\label{wilson_line_corr}
WL(R) \equiv \langle  \sum_{{\cal L}_{xy}}V({\cal{L}}_{xy}) \sum_{{\cal L}_{x^\prime y^\prime}}V({\cal{L}}_{x^{\prime}y^{\prime}})  \rangle -  \langle  V({\cal{L}}_{xy})  \rangle^2 \ ,
\end{eqnarray}
where the sums run over parallel Wilson lines placed on two parallel planes (the walls) separated by $R$ lattice spacings. 
${\cal{L}}_{xy}$ can be either a straight line or a $\sqcap$ line, as those described in Sec.2.2. These correlations give access to the mass of the Higgs field. 

\section{Numerical results}
\label{num_results}

\subsection{Monte Carlo updating}

The model has been numerically simulated on lattice volumes $L^3$.
Each updating step consisted in alternating blocks of five standard Metropolis hits for the gauge fields
with blocks of five standard Metropolis hits for the Higgs fields \cite{Metropolis}. Successive measurements of any observable were separated by $5-10$ such steps.
When we have studied transition lines towards the Higgs phase, which is presumed to behave like a spin-glass, a number of values of $\gamma$ for fixed $\beta$ have been gathered in one single simulation by use of the tempered Monte Carlo technique \cite{Marinari, Hukushima} in order to improve decorrelation efficiency. Moreover, the technique of replicas has been used to increase ergodicity, see for example \cite{castellani}. For given parameters $\beta$, $\gamma$, this trick consists in repeating the simulation $N_r$ times with different initial configurations and different random seeds,
and then average the $N_r$ results. In this fashion it is expected that a larger region of phase space is probed, even for poorly ergodic simulation algorithms.
Both techniques, tempered and replicas, are routine in standard simulations of spin-glass phases.

On the other hand, although in principle applying cluster algorithms \cite{Swendsen, Hoshen, Shiloach} for the Higgs fields would be feasible, we discarded that type of updating method
because during a few tests it proved to be scarcely ergodic. In particular it provided nonzero results for gauge-dependent functions, like the expectation value of gauge fields,
which is clearly inadequate.
Notice that this bad performance of cluster algorithms is in line with standard simulations of magnetic systems within their spin-glass phases, when the latter exist.

\subsection{The overlap operator}

\begin{figure}[H]
\centering
\includegraphics[width=0.65\linewidth,clip]{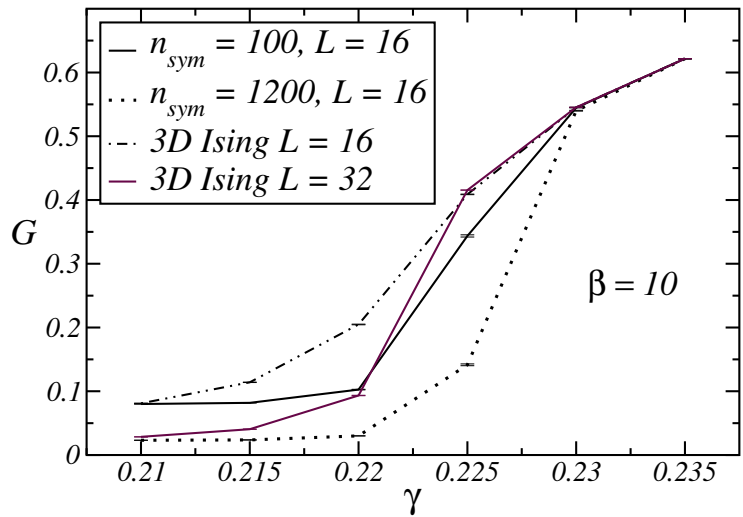}
\caption{Overlap operator at $\beta=10$ as a function of $\gamma$ for two values of ${n_{\rm sym}}$.}
\label{fig:overlap_vs_gamma_b10.0}
\end{figure}

\begin{figure}[H]
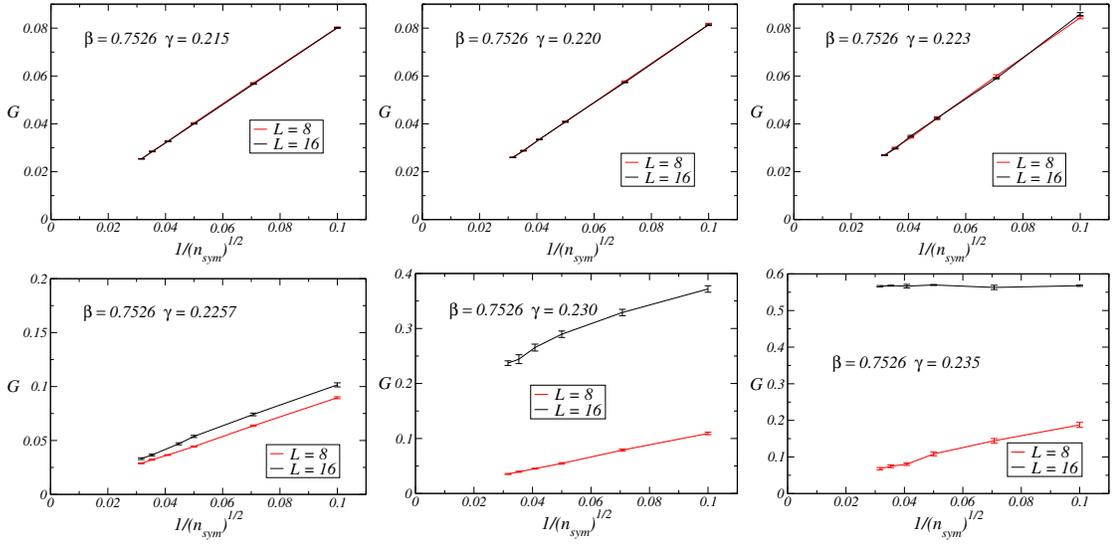

\centering
\includegraphics[width=0.32\linewidth,clip]{overlap_0.7526_0.215.eps}
\includegraphics[width=0.32\linewidth,clip]{overlap_0.7526_0.220.eps}
\includegraphics[width=0.32\linewidth,clip]{overlap_0.7526_0.223.eps}
\includegraphics[width=0.32\linewidth,clip]{overlap_0.7526_0.2257.eps}
\includegraphics[width=0.32\linewidth,clip]{overlap_0.7526_0.230.eps}
\includegraphics[width=0.32\linewidth,clip]{overlap_0.7526_0.235.eps}
\caption{Overlap operator in the vicinity of the multi-critical point at $\beta=0.7526$.}
\label{fig:overlap.b0.7526}
\end{figure}

\begin{figure}[H]
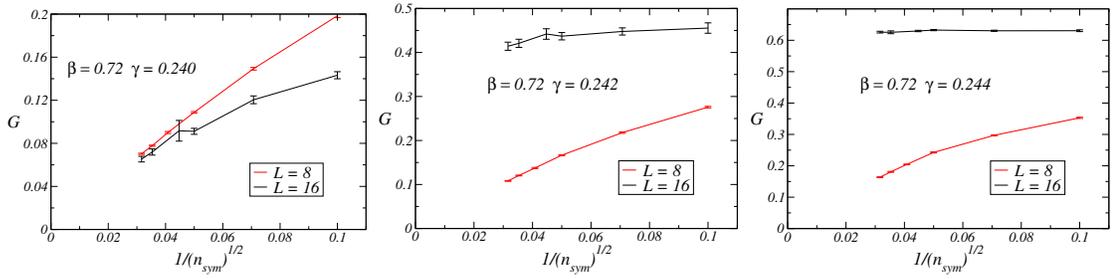

\centering
\includegraphics[width=0.32\linewidth,clip]{overlap_0.72_0.240.eps}
\includegraphics[width=0.32\linewidth,clip]{overlap_0.72_0.242.eps}
\includegraphics[width=0.32\linewidth,clip]{overlap_0.72_0.244.eps}
\caption{Overlap operator in the first order phase transition region near $\beta=0.72$.}
\label{fig:overlap.b0.72}
\end{figure}

\begin{figure}[H]
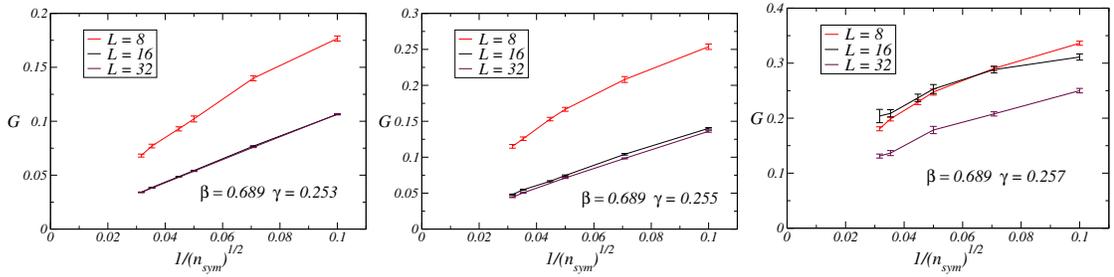

\centering
\includegraphics[width=0.32\linewidth,clip]{overlap_0.689_0.253.eps}
\includegraphics[width=0.32\linewidth,clip]{overlap_0.689_0.255.eps}
\includegraphics[width=0.32\linewidth,clip]{overlap_0.689_0.257.eps}
\caption{Overlap operator in the vicinity of the critical end point at $\beta=0.689$.}
\label{fig:overlap.b0.689}
\end{figure}

\begin{figure}[H]
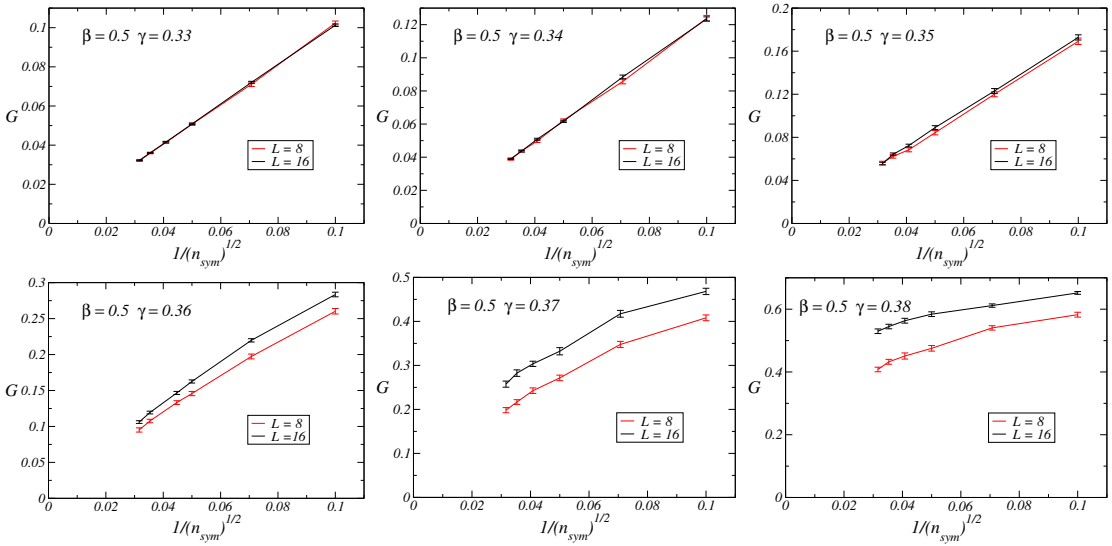

\centering
\includegraphics[width=0.32\linewidth,clip]{overlap_0.50_0.33.eps}
\includegraphics[width=0.32\linewidth,clip]{overlap_0.50_0.34.eps}
\includegraphics[width=0.32\linewidth,clip]{overlap_0.50_0.35.eps}
\includegraphics[width=0.32\linewidth,clip]{overlap_0.50_0.36.eps}
\includegraphics[width=0.32\linewidth,clip]{overlap_0.50_0.37.eps}
\includegraphics[width=0.32\linewidth,clip]{overlap_0.50_0.38.eps}
\caption{Overlap operator in the confinement-Higgs region at $\beta=0.5$.}
\label{fig:overlap.b0.50}
\end{figure}

\begin{figure}[H]
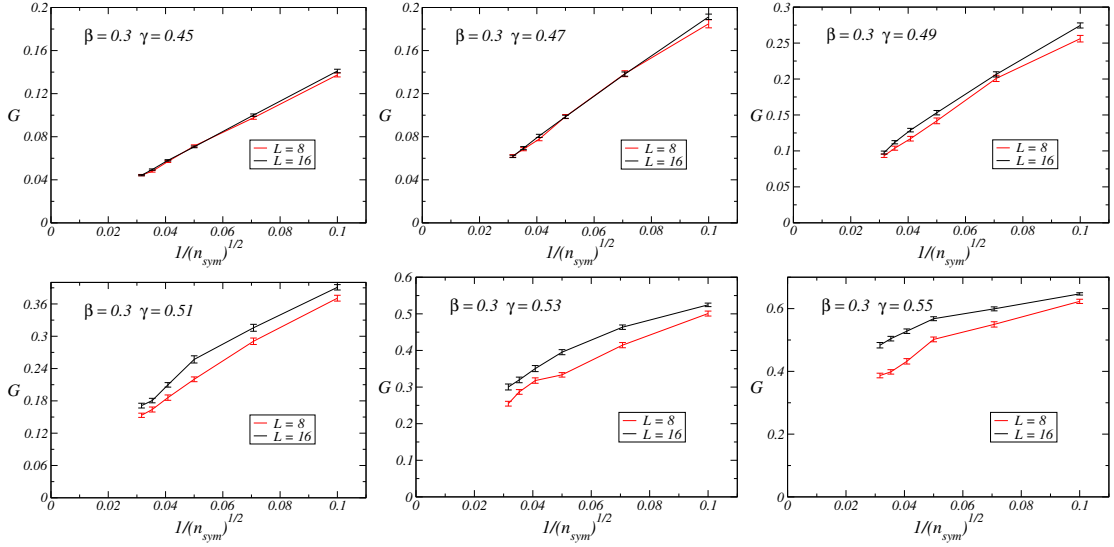

\centering
\includegraphics[width=0.32\linewidth,clip]{overlap1_00.30000_00.45000.eps}
\includegraphics[width=0.32\linewidth,clip]{overlap1_00.30000_00.47000.eps}
\includegraphics[width=0.32\linewidth,clip]{overlap1_00.30000_00.49000.eps}
\includegraphics[width=0.32\linewidth,clip]{overlap1_00.30000_00.51000.eps}
\includegraphics[width=0.32\linewidth,clip]{overlap1_00.30000_00.53000.eps}
\includegraphics[width=0.32\linewidth,clip]{overlap1_00.30000_00.55000.eps}
\caption{Overlap operator in the confinement-Higgs region at $\beta=0.3$.}
\label{fig:overlap.b0.30}
\end{figure}

Let us begin the discussion of the results from our Monte Carlo simulations with the analysis of the overlap operator.
Defined in (\ref{overlapoperator}), this operator has been measured for several values of $\beta,\gamma$ above and below the line separating the Higgs phase with other phases.
The results are shown in Figs.\ref{fig:overlap_vs_gamma_b10.0}-\ref{fig:overlap.b0.0}. For each considered value of $\beta$, the behavior of the overlap operator as a function of $1/\sqrt{n_{\rm sym}}$ is
shown for different $\gamma$ across the expected transition and for several lattice sizes. Each datum in these plots is the result of a few hundreds of Monte Carlo
measurements on equilibrium configurations combined with 200 replicas (thus, the full number of measurements was typically of the order of $4\times 10^4$).
Error bars were determined by jackknife within different blocking levels.

One observes that for each $\beta$ there is a critical value of $\gamma$ below which the overlap operator behaves linearly with
$1/\sqrt{n_{\rm sym}}$ and extrapolates to zero in the limit of infinite $n_{\rm sym}$; for larger values of $\gamma$ this feature is lost,
at least on the larger lattices. Here, the overlap operator takes on nonvanishing values.

We start by considering the large $\beta$ limit. In this limit the gauge
field becomes pure gauge and by a change of variables the theory can be reduced to the $3d$
Ising model. Thus, one expects that the overlap takes on a nonzero value at the critical point of the $3d$ Ising model $\gamma\approx0.2216$. Moreover,
the qualitative behavior of the overlap as a function of $\gamma$ should mimic the behavior of the magnetization of the Ising model. In order to check these features we simulated the model at a value of
$\beta$ as large as 10. The results are shown in Figs.\ref{fig:overlap_vs_gamma_b10.0}-\ref{fig:overlap.b10.0} and they confirm the expected behavior.

Next we studied the overlap in the region near the deconfinement-Higgs transition. Concretely for $\beta=1.0, 0.9,$ and $0.8$. In the language of spin-glasses this region corresponds to the ferromagnetic-glassy region, and the overlap operator should be able to distinguish the two phases. One expects that the overlap vanishes in the deconfinement (ferromagnetic) phase and gets a nonzero value in the Higgs (glassy) phase. Figs.\ref{fig:overlap.b1.0}-\ref{fig:overlap.b0.8} show that this is indeed the case and the critical points are compatible with the critical line of the Ising-like phase transition.

The behavior of the overlap near the MCP is shown in Fig.\ref{fig:overlap.b0.7526}. As expected, it is nonzero above the critical value $\gamma\approx 0.2258$.

Fig.\ref{fig:overlap.b0.72} demonstrates the effectiveness of the overlap operator to study the behavior of the model across the first order phase transition line at $\beta=0.72$. 
The critical point lies in the region $0.240\leq \gamma \leq 0.242$ and is compatible with the point on the self-dual line at $\gamma=0.2415$. Maybe naively, one could expect that the overlap continue to exhibit critical behavior along the self-dual line. However, this is not the case. Inspecting Fig.\ref{fig:overlap.b0.689}, which shows the overlap near the CEP, one concludes that the overlap is already nonzero when $\gamma=0.257$, while the self-dual point equals $0.2576$. This deviation from the self-dual line becomes larger and larger for decreasing $\beta$, a fact which is reflected in Figs.\ref{fig:overlap.b0.60}-\ref{fig:overlap.b0.0}. The approximate critical line above which the overlap is nonzero is shown in Fig.\ref{fig:phase_diagram_with_overlap}.

It is instructive to consider another limiting case, $\beta=0$. Here the partition function can be easily calculated analytically and the free energy
\begin{equation}
  \frac{1}{L^3}\ln Z=\ln\cosh\gamma
\end{equation}
does not exhibit any critical behavior. The full action has the form shown in Eq.(\ref{higgs_action}), which is called Edwards-Anderson model \cite{edwards_75}, and the random coupling takes the values $\pm 1$ with equal probability. Such a model is well known in the spin-glass theory and is called binary model. The transition to a spin-glass phase occurs in this model near the value $\gamma\approx 0.89$ \cite{ising_3dsg_review}. This value has to be confronted with the critical value $\gamma\in [0.74-0.78]$ which follows from analyzing Fig.\ref{fig:overlap.b0.0}. 

\begin{figure}[H]
\centering
\includegraphics[width=0.32\linewidth,clip]{overlap_10.00_0.210.eps}
\includegraphics[width=0.32\linewidth,clip]{overlap_10.00_0.215.eps}
\includegraphics[width=0.32\linewidth,clip]{overlap_10.00_0.220.eps}
\includegraphics[width=0.32\linewidth,clip]{overlap_10.00_0.225.eps}
\includegraphics[width=0.32\linewidth,clip]{overlap_10.00_0.230.eps}
\includegraphics[width=0.32\linewidth,clip]{overlap_10.00_0.235.eps}
\caption{Overlap operator at $\beta=10$.}
\label{fig:overlap.b10.0}
\end{figure}

\begin{figure}[H]
\centering
\includegraphics[width=0.32\linewidth,clip]{overlap1_01.00000_00.21000.eps}
\includegraphics[width=0.32\linewidth,clip]{overlap1_01.00000_00.22000.eps}
\includegraphics[width=0.32\linewidth,clip]{overlap1_01.00000_00.22500.eps}
\includegraphics[width=0.32\linewidth,clip]{overlap1_01.00000_00.23000.eps}
\includegraphics[width=0.32\linewidth,clip]{overlap1_01.00000_00.24000.eps}
\includegraphics[width=0.32\linewidth,clip]{overlap1_01.00000_00.25000.eps}
\caption{Overlap operator at $\beta=1$.}
\label{fig:overlap.b1.0}
\end{figure}

\begin{figure}[H]
\centering
\includegraphics[width=0.32\linewidth,clip]{overlap_0.90_0.210.eps}
\includegraphics[width=0.32\linewidth,clip]{overlap_0.90_0.220.eps}
\includegraphics[width=0.32\linewidth,clip]{overlap_0.90_0.225.eps}
\includegraphics[width=0.32\linewidth,clip]{overlap_0.90_0.230.eps}
\includegraphics[width=0.32\linewidth,clip]{overlap_0.90_0.240.eps}
\includegraphics[width=0.32\linewidth,clip]{overlap_0.90_0.250.eps}
\caption{Overlap operator at $\beta=0.9$.}
\label{fig:overlap.b0.9}
\end{figure}

\begin{figure}[H]
\centering
\includegraphics[width=0.32\linewidth,clip]{overlap_0.80_0.210.eps}
\includegraphics[width=0.32\linewidth,clip]{overlap_0.80_0.220.eps}
\includegraphics[width=0.32\linewidth,clip]{overlap_0.80_0.225.eps}
\includegraphics[width=0.32\linewidth,clip]{overlap_0.80_0.230.eps}
\includegraphics[width=0.32\linewidth,clip]{overlap_0.80_0.235.eps}
\includegraphics[width=0.32\linewidth,clip]{overlap_0.80_0.240.eps}
\caption{Overlap operator at $\beta=0.8$.}
\label{fig:overlap.b0.8}
\end{figure}

\begin{figure}[H]
\centering
\includegraphics[width=0.32\linewidth,clip]{overlap_0.60_0.29.eps}
\includegraphics[width=0.32\linewidth,clip]{overlap_0.60_0.30.eps}
\includegraphics[width=0.32\linewidth,clip]{overlap_0.60_0.31.eps}
\caption{Overlap operator in the confinement-Higgs region at $\beta=0.6$.}
\label{fig:overlap.b0.60}
\end{figure}

\subsection{Distance $r_{\alpha,\beta}$}

\begin{figure}[H]
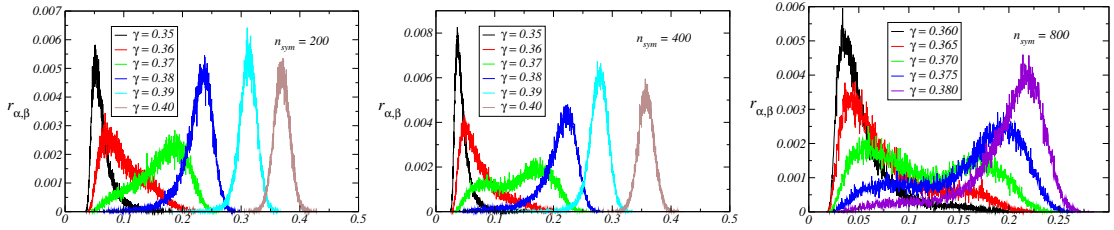

	\centering
	\includegraphics[width=0.32\linewidth,clip]{histo_b0.50_016_nsym200.eps}
	\includegraphics[width=0.32\linewidth,clip]{histo_b0.50_016_nsym400.eps}
	\includegraphics[width=0.32\linewidth,clip]{histo_b0.50_016_nsym800.eps}
	\caption{Histogram of distances between configurations in the confinement-Higgs region for different values of $n_{\rm sym}$ at $\beta=0.5$ on a $L=16$ lattice. The typical bin size is $10^{-4}$.}
	\label{fig:histo_b0.50}
\end{figure}

Now we discuss the results obtained with the operator (\ref{distance_operator}). The typical behavior of this operator in the confinement-Higgs region is displayed in Fig.\ref{fig:histo_b0.50} for $\beta=0.5$ at various values of $n_{\rm sym}$ on a $L=16$ lattice. One observes a change in the histogram of the operator near the critical point $\gamma\approx 0.37$. This value is compatible with the value where the overlap operator becomes nonzero in Fig.\ref{fig:overlap.b0.50}. The histogram shows a one-peak structure in the confinement and Higgs regions away from the critical point. Instead, in the vicinity of the critical point the histogram develops a two-peak structure as $n_{\rm sym}$ increases.  We have not measured this operator in the deconfinement-Higgs phase as the related simulations would require an extremely large number of replicas and, for large $n_{\rm sym}$, such simulations become hardly feasible, at least with our limited computer power.

\begin{figure}[H]
\centering
\includegraphics[width=0.32\linewidth,clip]{overlap_0.00_0.70.eps}
\includegraphics[width=0.32\linewidth,clip]{overlap_0.00_0.74.eps}
\includegraphics[width=0.32\linewidth,clip]{overlap_0.00_0.78.eps}
\includegraphics[width=0.32\linewidth,clip]{overlap_0.00_0.82.eps}
\includegraphics[width=0.32\linewidth,clip]{overlap_0.00_0.86.eps}
\includegraphics[width=0.32\linewidth,clip]{overlap_0.00_0.90.eps}
\caption{Overlap operator at $\beta=0$.}
\label{fig:overlap.b0.0}
\end{figure}

\subsection{The FM operator}

\begin{figure}[H]
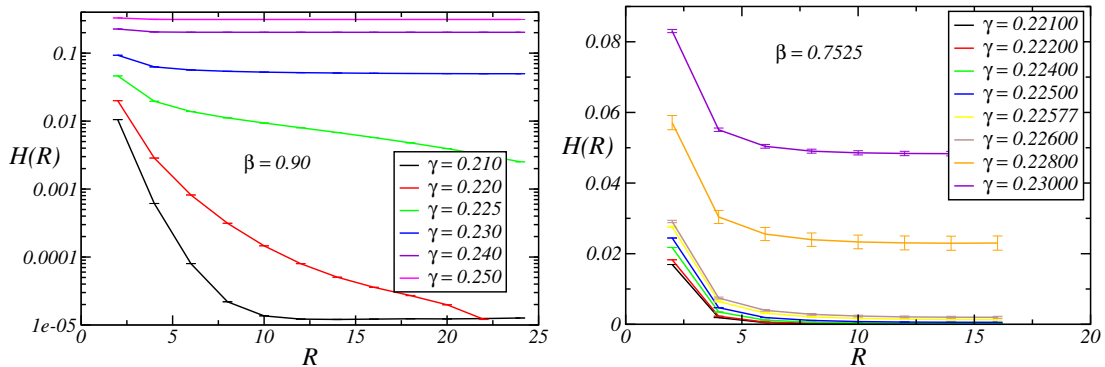

\centering
\includegraphics[width=0.48\linewidth,clip]{FM_048_00.90000_Fig15L.eps}
\includegraphics[width=0.48\linewidth,clip]{FM_032_0.7525_Fig15R.eps}
\caption{Left panel: FM operator at $\beta=0.90$ on a $L=48$ lattice.
Right panel: FM operator in the vicinity of the multi-critical point at $\beta=0.7525$ on a $L=32$ lattice.}
\label{fig:FM.b0.90_b0.7525}
\end{figure}

Here we present our results on the FM operator.
The FM operator can only distinguish deconfinement from confinement or Higgs phases. So it makes sense to simulate this operator in regions of the phase diagram where the system can be in the deconfinement phase. We have decided, therefore, to compute the FM operator at two points: 1) deeply in the deconfinement phase,  
$\beta=0.90$ and 2) in the vicinity of the MCP. Simulations were performed on  lattices $L=16, 32, 48, 64$ for $\beta=0.9$ and on the lattice $L=32$ for $\beta=0.7525$. In both cases, several $\gamma$ values were considered across the expected transition point. In reality, since the large distance limit cannot be attained numerically, we simulated the quantity $H(R)$ defined in (\ref{FMoperator_def}). In particular, the ratio $H(R)$ was determined  
for the square Wilson loop $R\times R$ and Wilson lines $R\times R/2$ for values of the distance $R$ ranging from~2 to $L/2$. Results are summarized in Fig.\ref{fig:FM.b0.90_b0.7525} and in Table 1.1. Each single datum in these plots is the result of $10^6$ Monte Carlo measurements on equilibrium configurations. Error bars were determined by the jackknife method with blocking.

The observed dependence on the distance can be summarized as follows: below a certain critical value of $\gamma$, the ratio drops rapidly towards zero for
increasing $R$, whereas at larger values of $\gamma$ it approaches a constant value, which increases with increasing $\gamma$. 
While the approach to a constant can be easily
identified, the approach to zero is a more delicate problem. In order to be sure
that the FM operator does vanish in the deconfinement phase, we performed simulations
on several lattices sizes $L=16, 32, 48, 64$ for three $\gamma$ values in the deconfinement
phase. Results obtained are shown on Fig.~\ref{fig:FM_b0.9_deconf}. These plots clearly
indicate that, within errors, the FM operator does vanish in the deconfinement phase as expected.
We did not study the scaling behavior in the vicinity of the critical point as it requires a huge numerical effort.

  \begin{center}
	\begin{tabular}{ |c||c|c|c|}
		\hline
		$R$ & $\gamma=0.21$ & $\gamma=0.22$ & $\gamma=0.225$ \\
		\hline
		2 &   $1.04970(26)\cdot10^{-2}$ &  $1.99443(63)\cdot10^{-2}$  &  $4.6162(18)\cdot10^{-2}$ \\
		\hline
		4 &   $6.0000(75)\cdot10^{-4}$ &  $2.8233(27)\cdot10^{-3}$  &  $1.9426(16)\cdot10^{-2}$ \\
		\hline
		6 &   $6.742(20)\cdot10^{-5}$  &  $7.817(16)\cdot10^{-4}$  &  $1.3657(14)\cdot10^{-2}$ \\
		\hline
		8 &   $9.846(94)\cdot10^{-6}$  &  $2.7844(96)\cdot10^{-4}$  &  $1.0958(13)\cdot10^{-2}$ \\
		\hline
		10 &   $1.639(33)\cdot10^{-6}$ &  $1.1288(52)\cdot10^{-4}$  &  $9.185(12)\cdot10^{-3}$ \\
		\hline
		12 &   $2.79(15)\cdot10^{-7}$ &  $4.897(34)\cdot10^{-5}$  &  $7.800(11)\cdot10^{-3}$ \\
		\hline
		14 &   $4.34(50)\cdot10^{-8}$ &  $2.204(50)\cdot10^{-5}$  &  $6.619(11)\cdot10^{-3}$ \\
		\hline
		16 &   $7.5(2.5)\cdot10^{-9}$ &  $9.91(16)\cdot10^{-6}$  &  $5.5755(91)\cdot10^{-3}$ \\
		\hline
		18 &   $1.9(1.2)\cdot10^{-9}$ &  $4.34(10)\cdot10^{-6}$  &  $4.6412(80)\cdot10^{-3}$ \\
		\hline
		20 &   $-$                                        &   $1.553(48)\cdot10^{-6}$  &  $3.8068(69)\cdot10^{-3}$ \\
		\hline
		22 &   $-$                                        &  $3.46(16)\cdot10^{-7}$  &  $3.0654(58)\cdot10^{-3}$ \\
		\hline
		24 &   $-$                                        &  $-$                                       &  $2.4151(46)\cdot10^{-3}$ \\
		\hline
	\end{tabular}
\end{center}
\vskip -0.2cm
\centerline{Table~1.1: FM data for $\beta=0.90$, at three values of $\gamma$ and several distances $R$ on $L=48$.}

\begin{figure}[H]
\centering
\includegraphics[width=0.32\linewidth,clip]{FM_00.90000_00.15000_Fig16L.eps}
\includegraphics[width=0.32\linewidth,clip]{FM_00.90000_00.18000_Fig16M.eps}
\includegraphics[width=0.32\linewidth,clip]{FM_00.90000_00.21000_Fig16R.eps}
\caption{FM operator at $\beta=0.9$ in the deconfinement phase. Data at different $L$ have been slightly shifted 
along the horizontal axix for clarity purposes.}
\label{fig:FM_b0.9_deconf}
\end{figure}

\subsection{Plaquette and Wilson line correlation functions}

We have considered three kinds of connected wall-wall correlators, according to the definitions given in Eqs.(\ref{plaq_plaq_corr}), (\ref{wilson_line_corr})
for a few values of $\beta$ and varying $\gamma$ across the expected transition line. $10^5 - 10^6$ data were typically collected for each pair of couplings $\beta,\gamma$, and
standard jackknife and blocking were used to estimate the uncertainties. In order to extract masses we performed a fit with a function of the form
\begin{equation}
C(R) = A \exp(-m R) / R^B \ , 
\label{correlation_fit}
\end{equation}
suitably made periodic to take into account boundary conditions. We used the
{\tt CERNLIB MINUIT} library to perform the fits, setting to $95\%$ the
confidence level in the estimation of the uncertainties on the fit parameters. 

\begin{figure}[H]
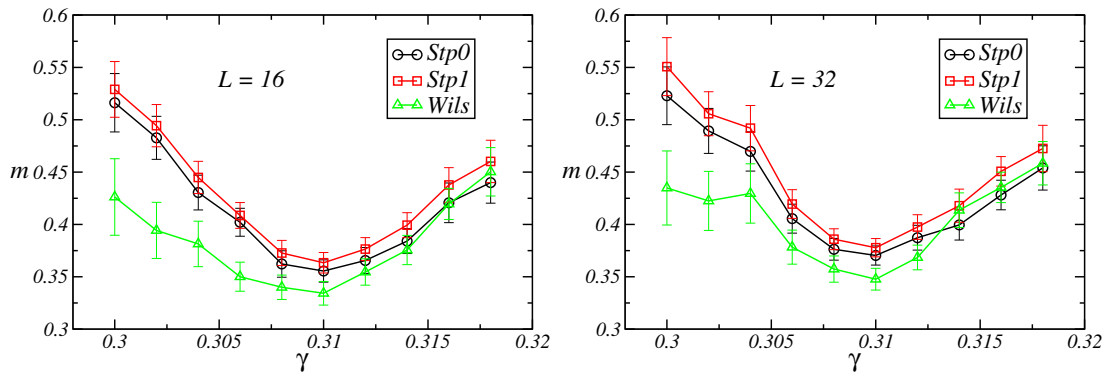

\centering
\includegraphics[width=0.48\linewidth,clip]{mass_b0.60_016.eps}
\includegraphics[width=0.48\linewidth,clip]{mass_b0.60_032.eps}
\caption{Masses near the confinement-Higgs transition, $\beta=0.6$. Left panel: $L=16$. Right panel: $L=32$}
\label{fig:mass_b0.60}
\end{figure}

We have computed the correlations and corresponding masses for $\beta=0.9, 0.6$, and $0.5$. 
Results for the masses in the confinement-Higgs region are shown in Figs.\ref{fig:mass_b0.60}-\ref{fig:mass_b0.50}. Here we have extracted masses from all three correlation functions. The nickname ``Stp0" denotes masses obtained from straight Wilson lines (with $T=0$ in the notation of subsections \ref{2.2FM} and \ref{2.3Corr}). The nickname ``Stp1" denotes masses obtained from staple-like Wilson lines, see again subsections \ref{2.2FM} and \ref{2.3Corr}. These masses are those determining the decay of the correlation between Higgs fields. Finally, ``Wils" denotes masses obtained from plaquette correlations which are governed by the mass of the gauge field. One observes that all three masses are not vanishing and develop the minimum close to the corresponding self-dual point.  For $\beta=0.6$ the position of the minimum is also close to the point where the overlap gets a nonzero value. Since all minima are stable and do not depend on the lattice size, one
may conjecture that the masses remain nonvanishing also in the thermodynamic limit. The masses extracted from straight and staple-like Wilson lines are compatible within errors. This agrees with the analytic result of Ref.\cite{FM_24}.

\begin{figure}[H]
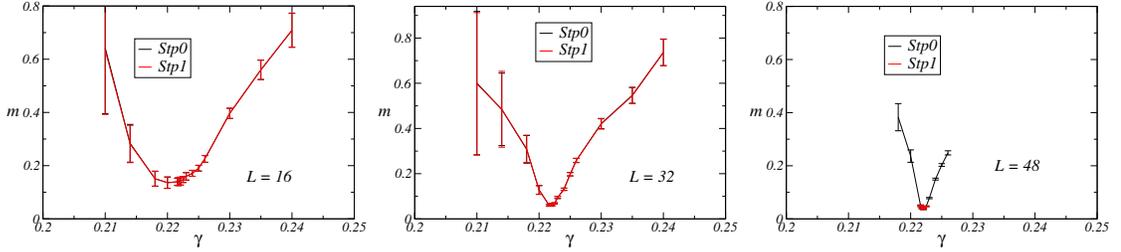

\centering
\includegraphics[width=0.32\linewidth,clip]{mass_b0.90_016_WL.eps}
\includegraphics[width=0.32\linewidth,clip]{mass_b0.90_032_WL.eps}
\includegraphics[width=0.32\linewidth,clip]{mass_b0.90_048_WL.eps}
\caption{Masses extracted from Wilson line correlations near the deconfinement-Higgs transition for $\beta=0.9$.
Left panel: $L=16$. Central panel: $L=32$. Right panel: $L=48$.}
\label{fig:mass_b0.90_wilsonlines}
\end{figure}

Results for the masses in the deconfinement-Higgs region are shown in Fig.\ref{fig:mass_b0.90_wilsonlines}. Here we computed only masses from Wilson line correlations. Again results ``Stp0" and ``Stp1" coincide within error bars. Unlike the transition from the confinement to the Higgs region, the minimum of the masses does depend on the lattice size and approaches zero with $L$ increasing. It is reasonable then to deduce that both masses vanish in the thermodynamic limit at the critical point. Modelling the scaling of the correlation length near the critical point with the form $\xi=1/m=a L$ where $a$ is a constant factor and $L$ the lattice size, and using data from the largest lattice $L=48$, we performed a fit to the functional form
\begin{equation}
	m = A |\gamma - \gamma_{\rm c} |^{\nu} + B 
	\label{mass_scale}
\end{equation}
where $A,B$ are constants. The results $\gamma_{\rm c}=0.222070(87)$, $\nu=1.92(49)$ and $\chi^2=0.14$ followed.

\begin{figure}[H]
\centering
\includegraphics[width=0.48\linewidth,clip]{mass_b0.50_016.eps}
\includegraphics[width=0.48\linewidth,clip]{mass_b0.50_032.eps}
\caption{Masses near the confinement-Higgs transition, $\beta=0.5$. Left panel: $L=16$. Right panel: $L=32$}
\label{fig:mass_b0.50}
\end{figure}

\section{Summary}

We studied by Monte Carlo simulation the $3d$ $Z(2)$ gauge-Higgs lattice model. 
Our prior objective was to test on a simple model the overlap operator proposed recently by Greensite and Matsuyama \cite{greensite_18_overlap}-\cite{greensite_22_overlap}. This operator was constructed by analogy of the gauge-Higgs action with spin-glasses \cite{edwards_75}. The pure gauge part of the full action acts as a probability distribution for random couplings in the Higgs part of the action which in the present case is nothing but the Ising action. This overlap operator was designed to distinguish confinement and Higgs phases. In $Z(2)$ gauge-Higgs LGT one has, in addition to confinement and Higgs phases, a deconfinement phase at large gauge and small Higgs coupling constants. Therefore, another goal is the present work was to check if the proposed operator is able to distinguish between deconfinement and Higgs phases.

In addition to the overlap operator we have also computed the more traditional nonlocal FM operator which is used to discriminate the deconfinement phase from the rest of the phase diagram.

Also a certain measure of the distance in phase space between configurations has been extracted following \cite{distance_config}.

\begin{figure}[htb]
\centering
\includegraphics[width=0.65\linewidth,clip]{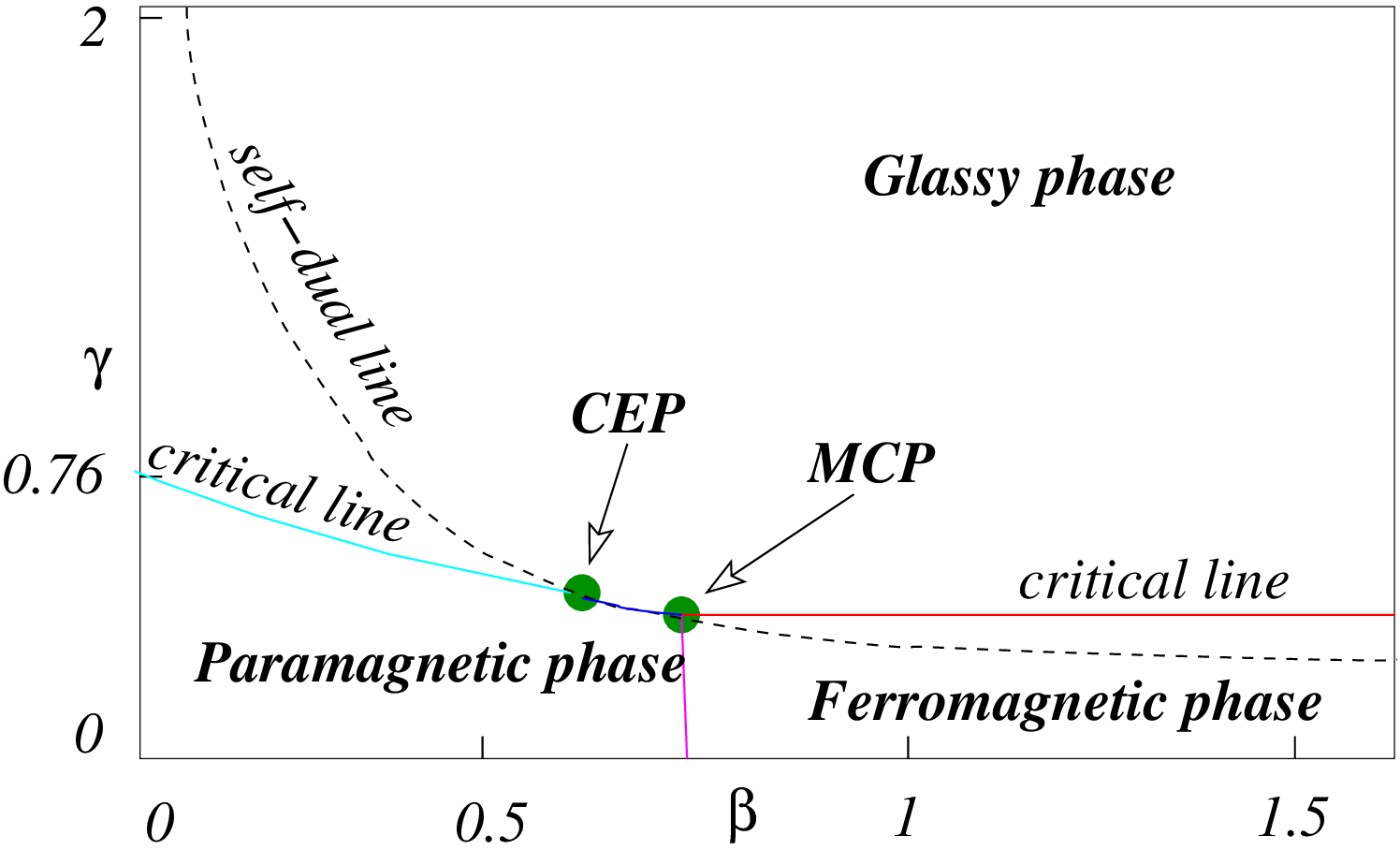}
\caption{Phase diagram of the $Z(2)$ gauge-Higgs LGT. Light blue line shows approximate critical line above which the overlap operator takes nonvanishing values.}
\label{fig:phase_diagram_with_overlap}
\end{figure}

Finally, we have studied plaquette and Wilson line correlation functions for a few values of the gauge coupling constant and extracted the gauge and Higgs masses of the theory. Our main results can be summarized as follows. 
\begin{itemize}
\item 
For all studied gauge couplings $\beta$, a value of $\gamma$ was found above which the overlap operator develops nonzero values. 
Our findings are summarized in Fig.\ref{fig:phase_diagram_with_overlap}.
The deconfinement-Higgs line (shown in red color) coincides with the expected
phase transition line. Starting from the MCP it continues along the self-dual
line until the CEP. For $\beta$ values smaller than the critical value at the CEP we observe a deviation from the self-dual line, and this deviation increases at smaller $\beta$ values (light blue line). Eventually, the light blue line 
reaches a point in the region $\gamma\in [0.74-0.78]$ for $\beta=0$. 
The general conclusion  is that the overlap operator does exhibit a nontrivial change when entering the Higgs phase either from the deconfinement phase or from the confinement phase and hence it can be used as order parameter to detect the Higgs phase. The principal differences between confinement and the Higgs phases were elaborated and explained in \cite{greensite_17_Sconf} for the $SU(2)$ gauge-Higgs model. These differences remain valid for the $Z(2)$ theory.  
\item 
The FM operator was computed near one of the deconfinement-Higgs critical points, 
$\beta=0.9$, and in the vicinity of the MCP. In the deconfinement phase the FM operator drops rapidly to a small constant for large distance. This constant is getting smaller with increasing lattice size. In the Higgs phase the FM operator gets a nonzero value in full agreement with theoretical predictions. 
\item 
The behavior of the gauge and scalar field masses was investigated for two values of the gauge coupling near the confinement-Higgs transition and for one value near the deconfinement-Higgs transition. In the former case it was found that both masses stay nonzero across the transition, whereas in the latter case the scalar field mass tends to vanish in the vicinity of the critical point. We think that these masses deserve a more thorough investigation which we plan for the future. 
\end{itemize}

{\bf Acknowledgments}. The authors acknowledge support from INFN/NPQCD project.
This work is partially supported by ICSC - National Centre for HPC, Big Data and Quantum Computing, funded by European Union -- NextGenerationEU.
Most numerical simulations have been performed on the CSN4 cluster of the Scientific Computing Center at INFN-Pisa.
O.B. thanks the Pisa section of INFN for the warm hospitality during a working stay.
B.A. is grateful to Juan Jos\'e Alonso for useful advice on simulation methods for spin-glasses.
B.A. and O.B. also thank Claudio Bonati, Massimo D'Elia and Claudio Bonanno for discussions.

\end{document}